# Drowning in Data : VO to the rescue

A. Lawrence

Institute for Astronomy, University of Edinburgh, Royal Observatory, Blackford Hill, Edinburgh EH9 3HJ, U.K.. (Scottish Universities Physics Alliance : SUPA)

**Summary**. Astronomical datasets are growing in size and diversity, posing severe technical problems. At the same time scientific goals increasingly require the analysis of very large amounts of data, and data from multiple archives. The Virtual Observatory (VO) initiative aims to make multiwavelength science and large database science as seamless as possible. It can be seen as the latest stage of a long term trend towards standardisation and collectivisation in astronomy. Within this inevitable trend, we can avoid the high energy style of building large fixed hierarchical teams, and keep the individualist style of astronomical research, if the VO is used to build a *facility class data infrastructure*. I describe how the VO works and how it may change in the Web 2.0 era.

## 1 Introduction

Modern astronomy has an embarassment of riches, but dealing with the volume and diversity of data presents severe technical challenges. At the same time, astronomers wish to address scientific problems with multi-wavelength data, and expect access to data to be as seamless and easy as shopping on the internet. The worldwide Virtual Observatory (VO) initiative is designed to address these problems. As we look at the key problems, we will gradually get a feeling for what the VO is and is not.

## 2 The Data Deluge

Many science goals in modern astronomy require very large datasets for their solution. Sometimes this is because large numbers of objects are needed to obtain the required statistical accuracy. For example, measuring the power spectrum of galaxy clustering, and finding small wiggles such as those due to baryon-acoustic oscillations, requires ~1% accuracy in many successive bins. Likewise, mapping dark matter by its weak lensing effect requires measuring the shapes of many faint galaxies in every one of a grid of many cells, in order to see a weak non-random tendency to line up. Other times, the necessity for large surveys is because intrinsically large objects are being studied - for example looking for star streams across the Milky Way which are a fossil record of its merger history. Finally, and increasingly fashionably, astronomers look for very rare, one in a billion objects - z=7 quasars, the Near Earth Object which will destroy the Earth, free floating planets. (An example is shown in Fig. 1).

As well as sheer data volume, modern astronomy often needs data intensive computing. Many of the calculations one wants to do with large datasets scale as $N^2$; we want to process large amounts of data in real time to spot transients such as Gamma Ray Bursts and alert other astronomers; and even telescope operations may soon involve supercomputing, for example to calculate the optical corrections needed to correct for the atmosphere in multi-conjugate adaptive optics.

So just how big is a sky survey ?  If you map the sky in one waveband with pixels 0.1 arcsec across, encoding the brightness at each spot with 16 bit accuracy, that makes a 100TB dataset.  Processing this pixel map to make an object catalogue typically makes a database ten times smaller than that. Modern sky survey catalogues (SDSS, UKIDSS) have of the order one billion objects. Then of course one can imagine repeating that for many wavelengths, and repeating again and again to sweep the sky for transients. Right now the UKIDSS survey (see below) is producing 20TB/yr; during 2009 VISTA should start producing 100TB/yr; by 2015 LSST will produce 5PB/yr; and by 2020, SKA will produce 100PB/yr.

If you want to get a concrete feel for what a big database is really like, try playing with the zoomable UKIDSS Galactic Plane Mosaic, at http://surveys.roe.ac.uk:8080/wsa/gps_mosaic.jsp.

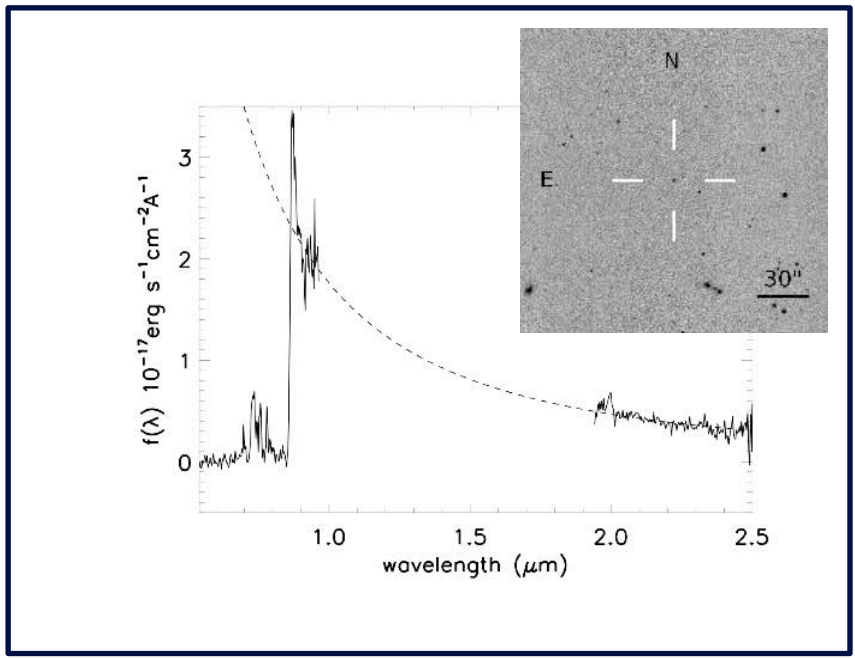

**Fig. 1** An example of large database science : infrared spectrum of the z=6.13 quasar ULAS J1319+0950 from Mortlock *et al* 2008. This work aims at finding very high redshift quasars by combining IR (UKIDSS) and optical (SDSS) data, looking for a rare combination of colours. Mortlock *et al* found four quasars at z~6 over an area of 900 square degrees, examining many millions of objects.

Thanks to Moore's Law, these data rates are not a fundamental technical problem; as computers get bigger and faster we will be able to store and process these datasets. However some things don't scale with Moore's Law. The first is the number of skilled workers per project; organising and operating the data processing required before astronomers can actually do science with the data gets ever harder. The second thing that is not growing with Moore's law is I/O, i.e. the speed with data can be moved between disk and CPU. The third thing is real-world end-user bandwidth. Your University may be connected to a Gbps backbone, but when you try FTPing something to your laptop, what you get is a thousand times slower. (This is the famous "last mile problem".) Using a typical PC over the network, searching a big modern database, or downloading such a database (even assuming you have somewhere to put it) could take all week.

While individual users are blocked by these problems, a good modern data centre can fix them without too much difficulty - they will set up a dedicated database server with multiple CPUs, many disks hanging off each motherboard, and an intelligent indexing system. If they also offer searching and other kinds of data analysis online and with a good interface, then you don't download the data, only your results. The data centre provides a *service*.

We have arrived at our first lesson about what the VO must be like. We are moving into a remote service economy : the motto is "*shift the results not the data*".  The VO is not a giant warehouse. Its not a hierarchical pipeline, like the LHC datagrid built by the particle physicists. Its not a democratic peer-to-peer system like Napster or SETI online. Its a small set of professional *service centres*, and a large population of *end-users*. In fact, its pretty much like shopping.

## 3 UKIRT Infrared Deep Sky Survey (UKIDSS)

The UKIRT Infrared Deep Sky Survey (UKIDSS; Lawrence *et al* 2007) is a good example of modern issues. (And a project for which I happen to be PI of course ...) It is a survey being carried out at near-IR wavelengths (ZYJHK) using the Wide Field Camera (WFCAM) on UKIRT. Unlike the 2MASS survey, it does not cover the whole sky, but it is much deeper. It is actually a portfolio of five subsurveys. (See Fig. 2.) Three of the surveys are wide area "shallow" surveys, covering approximately 7000 square degrees to a depth of around K=18.4 (Vega magnitude). Then there is a Deep Extragalactic Survey covering 35 square degrees to K=21, and an Ultra Deep Survey covering 0.7 square degrees to K=23. The survey started in 2005, takes around half the UKIRT time, and should complete by about 2011-12. More information is available at the UKIDSS web site, http://www.ukidss.org.

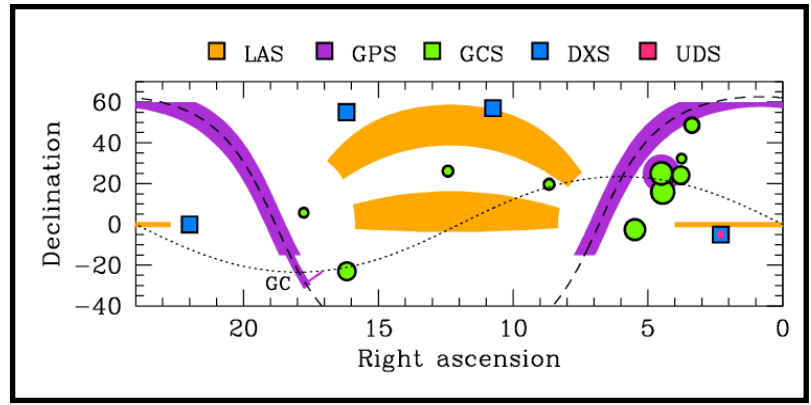

**Fig. 2** Sky Coverage of UKIDSS.

UKIDSS produces 100GB of new data every night. These data are sent to Cambridge for pipeline processing (Irwin *et al* 2009), and then up to Edinburgh for final processing and ingestion into the WFCAM Science Archive (WSA : Hambly *et al* 2008). The data can be accessed through the WSA at http://surveys.roe.ac.uk/wsa. To keep up with the dataflow, the Cambridge-Edinburgh internet link uses dedicated end-machines, carefully tuned TCP buffers, and multiple parallel streams. The WSA is growing at 20TB/year, and has over a billion objects in its database tables. Screenshots of the WSA in action are shown in Fig. 3.

The term "science archive" is meant to show that this is not just a repository, but a live resource used to actually do the science. Just as with the Sloan Digital Sky Survey (SDSS; York *et al* 2000), which pioneered this approach, the WSA consists of source catalogues held in a structured relational database. Remote users submit queries in Structured Query Language (SQL), along the lines of "give me a list of things in this piece of sky brighter than A with colour redder than B, that look starlike, are not blended, and have quality flag better than C" etc. This online service enables astronomers to do the kind of big-sample science described in Section 1 - for example, measuring the clustering of different types of galaxy, and finding rare objects. Through the WSA, astronomers have already found several z=6 quasars, and the coolest known brown dwarf.

**Fig. 3.** Screenshots from the WFCAM Science Archive (WSA), the web based interface to UKIDSS.

This brings us to VO lesson number two. More and more science is performed using on-line archives, with many different experiments performed on the same datasets. The motto here is "*the archive is the sky*".

## 4 The Tower of Babel

Although the sheer number of bits in astronomical databases produces problems, the real problem is the number of different archives. Every day, astronomers still get science out of legacy datasets like the Schmidt plates, IRAS, and Hipparcos. Meanwhile they get even more out of current hits like VLT, SDSS, 2MASS, HST, Chandra, XMM, WMAP, UKIDSS, and Fermi. We are keenly expecting more key datasets from VISTA, Planck, Herschel, LOFAR, and not too far away, ALMA and JWST. We are crossing our fingers for LSST, ELT, Lisa, Darwin, SKA, IXO, etc. This is far from a complete list of course.

This rich suite of complementary facilities and missions is a fantastic opportunity. We really do live in a golden age. But as we move further into the online access era, dealing with different data formats, access modes, user interfaces, and password systems etc, is maddening. The key issue then is *archive interoperability*. We need *standards*, and a *transparent data infrastructure*. Putting that into English, if everybody uses the same screw threads, the archives have nice fat pipes between them, applications writers know how to suck things down the pipes, and so on, things could be much much easier. So thats VO lesson number three : "*lets all use the same screwthreads*".

There is a nice way to think of this. The beautiful thing about the Web is its transparency. When you click on links and jump around, although those documents are in lots of different cities, they feel as if they are just *inside your PC*. The aim of the VO is to achieve the same feeling for data. All the world's data should feel as if its right there inside your own computer, and you just get stuck in and explore it. So thats lesson number four : "*The VO feels like its inside your own computer*".

Easier said than done of course ...The key step towards making this real has been the creation of an organisation for debating, writing, and agreeing standards - the International Virtual Observatory Alliance (IVOA : see http://www.ivoa.net. ). In the last few years, this body has agreed standards on data exchange formats, resource and service metadata, data access protocols, table column semantics, software component interfaces, virtual storage addressing, and more. As these standards are agreed, they are already being deployed by data centres and tools writers, but also passed on to the IAU for formal ratification.

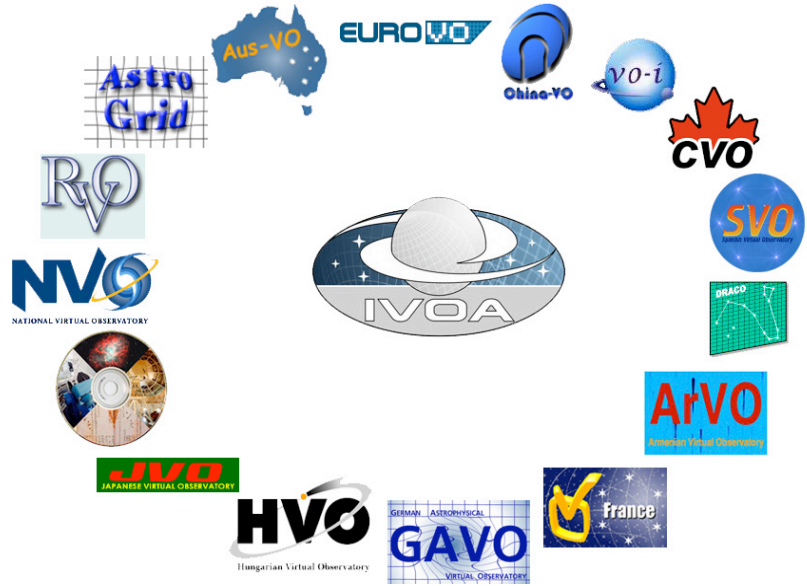

**Fig. 4** Logos of the worldwide projects making up IVOA.

We can now sum up what the VO *is* as opposed to what it *is not*. It is an evolving set of *software standards*; it is those data collections round the world that *follow the rules*; and it is an ever growing family of software tools that *understand the rules* and can do stuff with the data in the club. There is no central VO-command. There is no big VO-warehouse. The VO is not a *thing* at all; its a *way of life*.

## 5 We are the Borg

The worldwide VO initiative began around 2001, with major projects starting in the USA, the UK, and Europe-wide, which have been gradually joined by more VO projects all over the world. However, before 2001 there were previous attempts at projects with a similar vision that were just a little before their time. The NASA ADS system with its organised searchable online literature collection is the (very successful) remnant of an originally much more ambitious project. The resources collected and made available by CDS in Strasbourg, and by the HEASARC Browse system, were already proto-VOs.

The reason for the rapid technical progress since 2001 has been partly because of the determined re-start; but probably mostly because the time was right. Internet technology has progressed to a point where the necessary basic technology is openly available - notably XML, web services, and the increasing commoditisation (and cheapness) of relational database technology. This means that astronomers have only had to build an astronomy-specific layer, not to build an original technology.

Stepping back and looking at astronomy over the last few decades, it seems clear that actually the VO is simply the latest stage in a steady trend towards more standardisation, and more collectivisation. Let us look first at standardisation.

The first key step was the development of what was known in the UK as "common user instrumentation" and in the USA as "facility class instrumentation". The Anglo-Australian Telescope was particularly ahead of the game here. Rather than some University research group lashing together their own spectrograph, driving it up the mountain, and nursing it through an observing run, the Observatory builds a robust device that sits permanently on the telescope, and that any competent astronomer can turn up and *drive*. This is not a trivial step, as the standards of engineering, reliability, and documentation required are much greater - but the results have liberated huge amounts of cost effective science.

The next step was the standardisation of data formats - FITS, NDF, etc. Rapidly following on this was the production of "facility class" data reduction software - IRAF, MIDAS, Starlink, AIPS etc. Just like with the instruments, rather than getting your grad students to lash up amateur data reduction code, you learn to drive the standard packages, which understand those standard data formats. The VO is the next step in that process - standardising data access methods, data exchange formats, and metadata describing what a data resource contains and what you can do with it. This means that you can automate many things that previously needed to be done by hand. Finally, a logical next step is the standardisation of *data analysis* tools, as opposed to data reduction tools. This has already begun, with the now typical practice of a data centre offering scientifically meaningful SQL queries as a service, but will go further - press this button to Fourier Transform your dataset. One might refer to this trend as VO++.

These examples of increasing standardisation also have aspects of collectivisation, as they have happened through the collec-tive desire of the community to organise itself this way. No single astronomer or research group is likely anymore to build their own infrared camera, or construct a data reduction suite single handed. Which of these things even happens at all is decided communally through the peer review process and the long and broad debates around that. In recent years, even the *collection of data* has been collectivised. Although SDSS started out as a small team, it ended up as a huge professionalised project.

There is a puzzle here. The idea of VO services seems to make astronomy more like shopping, but on the other hand this process of collectivisation would seem to be turning us into the Borg, assimilated into an anonymous collective. Which wins ? Are we the Borg, or are we Happy Shoppers ? The answer is that we have a choice.

In the process of the drive towards big experiments and collectivisation, we are following behind the particle physicists. Their solution has been to make large coherent teams centred around specific projects, that take responsibility for the complete end-to-end chain - design, construction, data analysis, science. These teams have their own internal rules, and every paper is by Aardvark *et al*. By contrast, astronomy is notoriously individualist. If you are a smart postdoc you don't need to ask the permission of some Big Prof or Project Leader - you write a telescope proposal, get your three nights, write the paper with two other chums, get famous. This is possible because the construction of big telescopes and detectors is not just *collectivised*, its *professionalised*. People who specialise in designing and building the kit hand over the results of their labours as *open facilities*; the exploitation is decoupled.

So as astronomical surveys become big projects, and as the technology of data access and analysis becomes standardised, the way to pull off the same trick of empowering the smart postdocs is to think of those surveys and data centres as likewise being open facilities, constructed to a professional standard for others to use. In other words what we want is a *facility class data infrastructure*. Thats a pretty good definition of the VO at its most general.

## 6 Web 2.0 Astronomy

Is there a danger that the VO is already behind the times, being an old style Web project, as opposed to a trendy young Web 2-ish kind of thing ? Lets take a look at what this could mean.

The key Web metaphor is that of transparency. After years of technical development of the internet, there was a key point where the invention of HTML and graphical web browsers suddenly made documents worldwide trivially visible, and the whole wonderful world of web-surfing took off. As I have already explained, a goal of the VO is to achieve the same feeling for data. However, in the creation of the World Wide Web, there was initially a clear distinction between *creators* and *readers*, and likewise between *clients* and *servers*. Anybody could install Mosaic and click on those links; but setting up an Apache web server was the sort of thing only sys admins could do. As a result, content was largely published by organisations rather than individuals, who just passively browsed the content.

By contrast, the new "Web 2.0" style is interactive, participatory, and democratic. Users don't just read stuff. They create it (blogging); they adjust it (wikis); they vote on it (Digg etc); and they structure the metadata (tagging). It all has the feeling of content and meaning evolving spontaneously. By contrast, the VO world perhaps seems rigid - life is dominated by big missions and data centres; the IVOA dictates the standards and you must obey. Can the VO just *emerge* ?

There are both opportunities and dangers here. Firstly, the people-power aspect of Web 2.0 is something of an illusion. All the interactivity relies on a background infrastructure provided by commercial corporations. You can only write your blog on BlogSpot because Google is running special software on huge server farms that they allow you to use. You can't change that software anytime you want to, and it could get in principle get switched off. For science, we really want open standards and a genuinely public infrastructure. The next point is that it is not at all obvious that spontaneous user-driven tagging and so forth is capable of producing a reliable metadata structure for something as technically complex and specific as astronomical data access.

Nonetheless it does seem attractive to look for what the elements of "VO 2.0" might be. I can see four growth areas.

(1) <u>Astronomical tools</u>. So far VO tools have been written by pros within VO projects. But as the infrastructure matures, and Application Programming Interfaces (APIs) develop, it is becoming easier to write astronomical application tools that string together VO services and do other cool stuff. This is the key growth area.

(2) <u>Annotation</u>. The next obvious area is annotation - adding your own comments and tags to resources that you find and bookmark. You can already do this with AstroGrid's VODesktop application.

(3) <u>Data sharing</u>. Key advances made by IVOA include agreements on how to specify the location of remote storage, and how to specify your identity. Now these protocols are in place, it will be increasingly easy to swap and share data, saved queries, workflows etc. ("Hey Jean, that query worked well. I save the result in VOSpace and flagged it for your access, take a look".)

(4) <u>Free market standards</u>. Mostly, defining standards formally in the IVOA, and then implementing them, will remain the sane thing to do. However there may be some areas where it could make sense for working practices to evolve naturally and then be sanctioned post-hoc as de facto standards.  One of the most interesting possibilities is in data models. This is an area of the VO that is of great importance but is making slow progress because it is so hard to agree on relatively arbitrary details. Perhaps the answer is to allow anybody to *publish* a data model, and then each dataset can refer to a chosen data model. ("The structure of my catalogue follows the data model at this URL, so if your query works on that, it will work on mine."). The best data models will naturally emerge.

**7 The Virtual Observatory in action.**

The conference talk that this article results from concluded with a live demo. Here in print we will achieve a similar result by providing some links to working software. This is of course biased towards work I have been personally involved with.

Availability of large surveys through web pages interfacing to relational database management systems is becoming more common. The prime example is the SDSS Sky Server, at http://cas.sdss.org/astrodr7/en/. (Thakar et al 2008). Here you can do simple region queries,  or submit full-blown SQL queries. The SDSS web site also has  some excellent online browse-and-play tools,     such     as     the     Navigator,     at

http://cas.sdss.org/astrodr7/en/tools/chart/navi.asp. The new kid on the block is the WFCAM Science Archive, at http://surveys.roe.ac.uk/wsa/, which is where you get the UKIDSS data. As well as a fill-in-the-boxes interface and a free-form SQL interface, it has a marvelous zoomable mosaic of the Galactic Plane, at http://surveys.roe.ac.uk:8080/wsa/gps_mosaic.jsp.

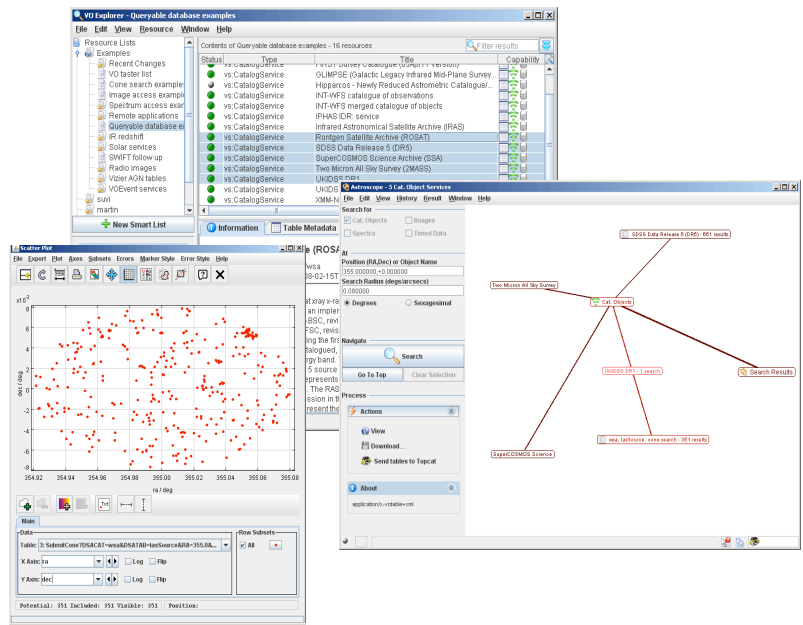

**Fig. 5** Screenshots from use of AstroGrid tools, shwoing how one can query several resources at once. The upper figure shows VOExplorer, a tool within VODesktop used to browse and search the Registry (yellow pages). In this example several resources are selected (highlighted in blue), and then a simple cone-search run at a chosen position, using the Astroscope tool (lower right). Finally, the tables resulting from querying those resources are piped to the Topcat tool (lower left) for analysis.

But now you can access the same data through VO tools, at the same time as querying other resources. Some VO tools operate as services run through web pages. The US-NVO project has a series of tools which work this way, which you can find through starting at the NVO web page, http://www.us-vo.org/. Likewise, the Japanese VO project operates as web-based portal, at http://jvo.nao.ac.jp/.

By contrast, the approach within Europe has mostly been to provide Java applications. These need a few simple installation steps, but provide a much richer and more flexible user interface. A good list of useful VO tools is provided at the Euro-VO tools page, http://www.euro-vo.org/pub/fc/software.html. A key aspect of these tools is that they speak to each other, so that if for example you have used found an image by querying the registry, you can send it directly to a compatible image browser, without an intermediate download, startup, and reload sequence.

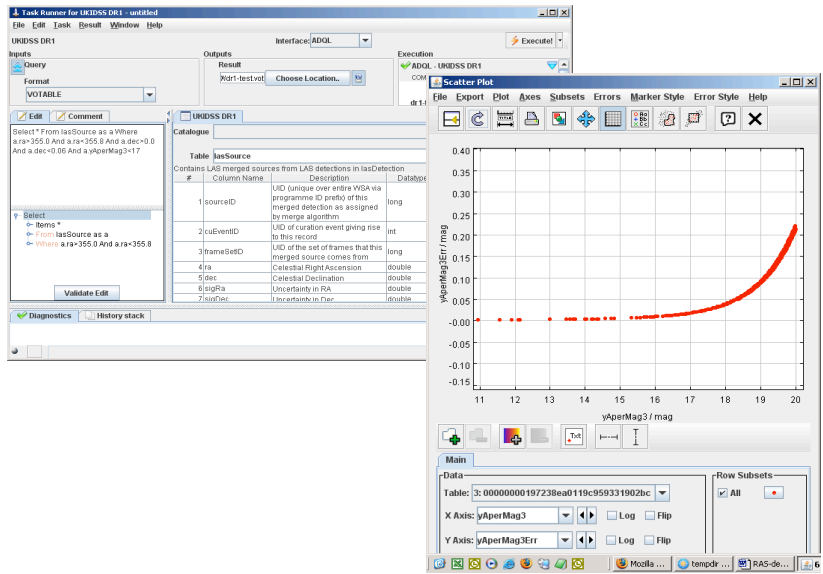

**Fig. 6.** More AstroGrid tools screenshots, this time showing the use of the generic Query Builder. If the selected resource is a database with SQL acces (rather than just a simple catalogue) then the database table names and column names appear in the registry entry put there by the data provider. The user can therefore construct an SQL query to send to the database, using the same tool whatever the database, rather than many different web page interfaces. As before, the results can be piped directly to Topcat for analysis.

In the UK AstroGrid project, we aim to provide a small but complete suite of tools that do what you need. These can be found at http://www.astrogrid.org/. The central tool is VODesktop. This allows you to browse, search, and bookmark entries in the Registry (Yellow Pages); send queries to selected resources; send results to other applications for analysis; and read and write from VOSpace. here "query" could mean a simple cone-search, or constructing a full-blown SQL query, or entering parameter values to run an application like HyperZ running remotely. The second key tool is Topcat, which is all about astronomical tables - browsing them, manipulating themm, and plotting them. The third key tool (from our French CDS colleagues) is Aladin, which is used for examining and analysing images. Finally, there is a Python package, which allows you to write Python scripts which do similar things - search the registry, download data , etc. This is harder than the GUI applications at first but of course much more powerful, as you can automate repetitive things, and mix the VO service calls with other calculations etc.

**8 Conclusions.**

After some years of conceptual and technical development, Virtual Observatory infrastructure and tools are now in a working but not quite mature state. The VO is already fit for use both by data providers and by data consumers, but is likely to evolve further as a new generation of astronomers, used to Web 2.0 technologies, adjust it to their liking. As a community, astronomers can develop their data infrastructure in a way that preserves their traditional individualism, even whilst carrying out big projects in a collective and professional manner.